\documentclass{ws-procs975x65}
\begin{document}

\title{The Complexity of Biological Ageing}

\author{Dietrich Stauffer} 

\address{Institute for Theoretical Physics, Cologne University, D-50923 K\"oln, Euroland}

\maketitle

\abstracts{
The present review deals with the computer simulation of biological
ageing as well as its demographic consequences for industrialized societies.}

% Keywords: Monte Carlo simulation, Penna model, sex, retirement.

\section{Introduction}
Life usually ends with death, and ageing is defined here by the increase of 
the mortality rate with increasing age. Merryl Streep and Goldie Hawn showed
in the movie ``Death becomes her'' the consequences of an elixir giving us
eternal life. The present review instead deals with the consequences present
foreseeable trends have on the demography of developed countries, and with
the biological reasons of ageing.

\begin{figure}[t]
%\figurebox{22pc}{15pc}{}
%\epsfbox{vancouver1.eps}
\includegraphics[angle=-90,scale=0.5]{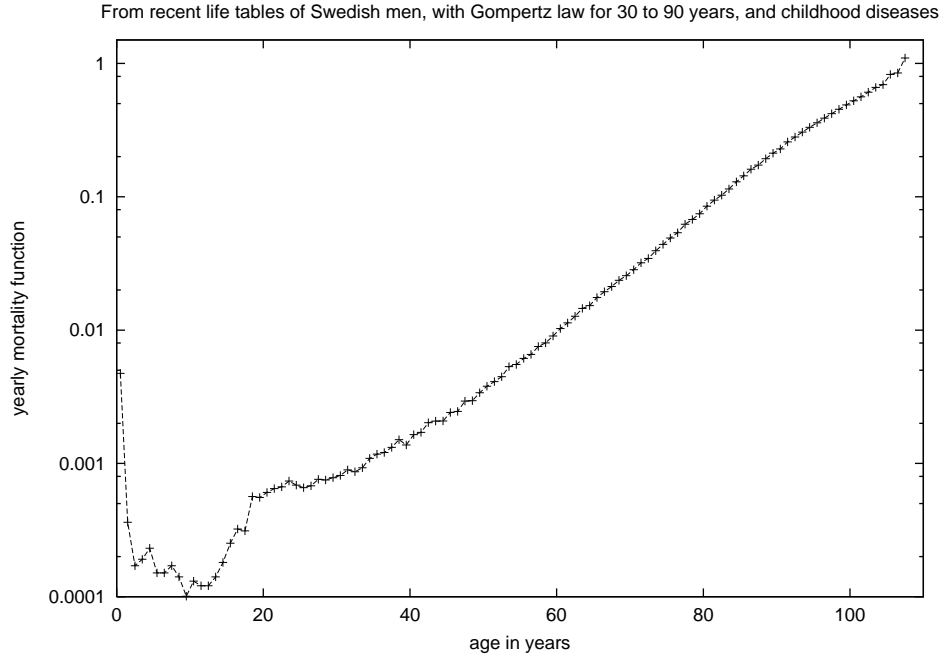}
\caption{Gompertz law ad middle age, childhood deviations, and possible downward
deviation from Gompertz law for centenarians among Swedish men. From life
tables 1993-1997 of the central statistical bureau in Sweden:
Ewa.Eriksson@scb.se (1999).}
\end{figure}

\begin{figure}[t]
%\figurebox{22pc}{15pc}{}
%\epsfbox{vancouver2.eps}
\includegraphics[angle=-90,scale=0.5]{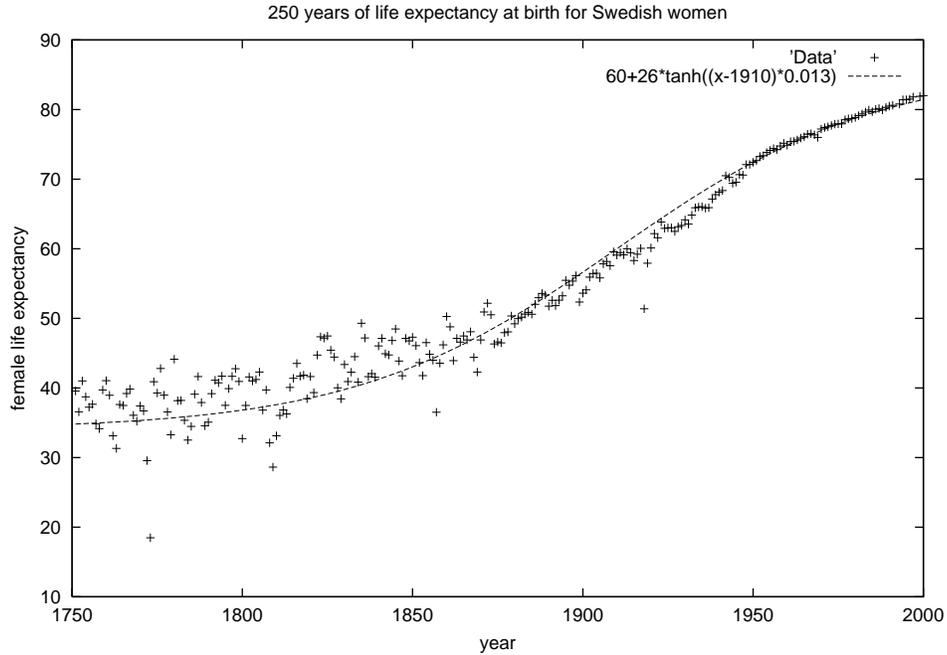}
\caption{Life expectancy for Swedish women for 250 calendar years. From 
Wilmoth's Berkeley Mortality Data Base.}
\end{figure}

\begin{figure}[t]
%\figurebox{22pc}{15pc}{}
%\epsfbox{vancouver3.eps}
\includegraphics[angle=-90,scale=0.5]{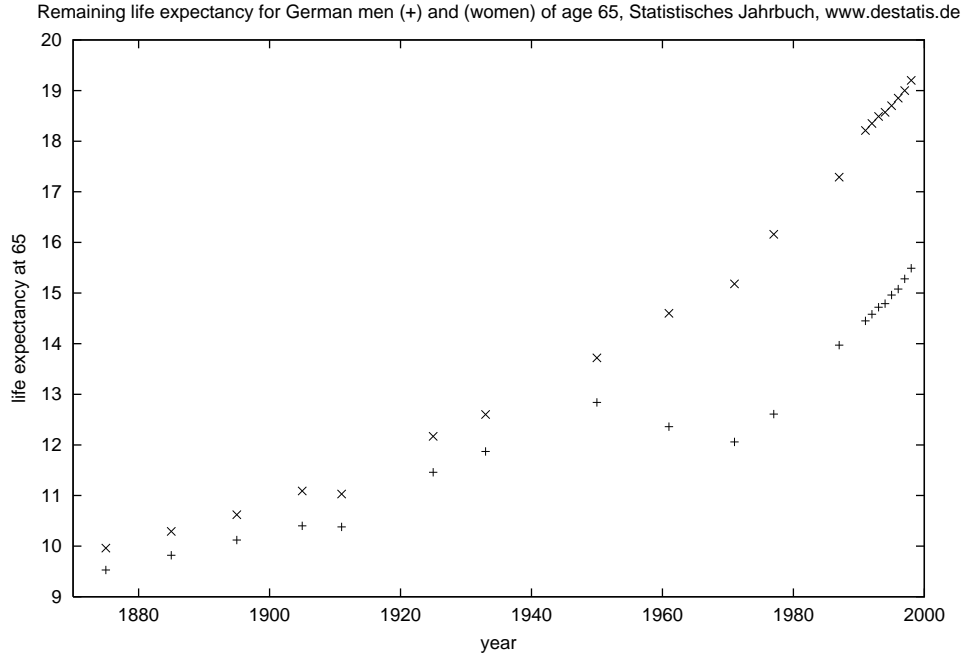}
\caption{Remaining life expectancy at the German legal retirement age of 65.}
\end{figure}

\begin{figure}[t]
%\figurebox{22pc}{15pc}{}
%\epsfbox{vancouver4.eps}
\includegraphics[angle=-90,scale=0.5]{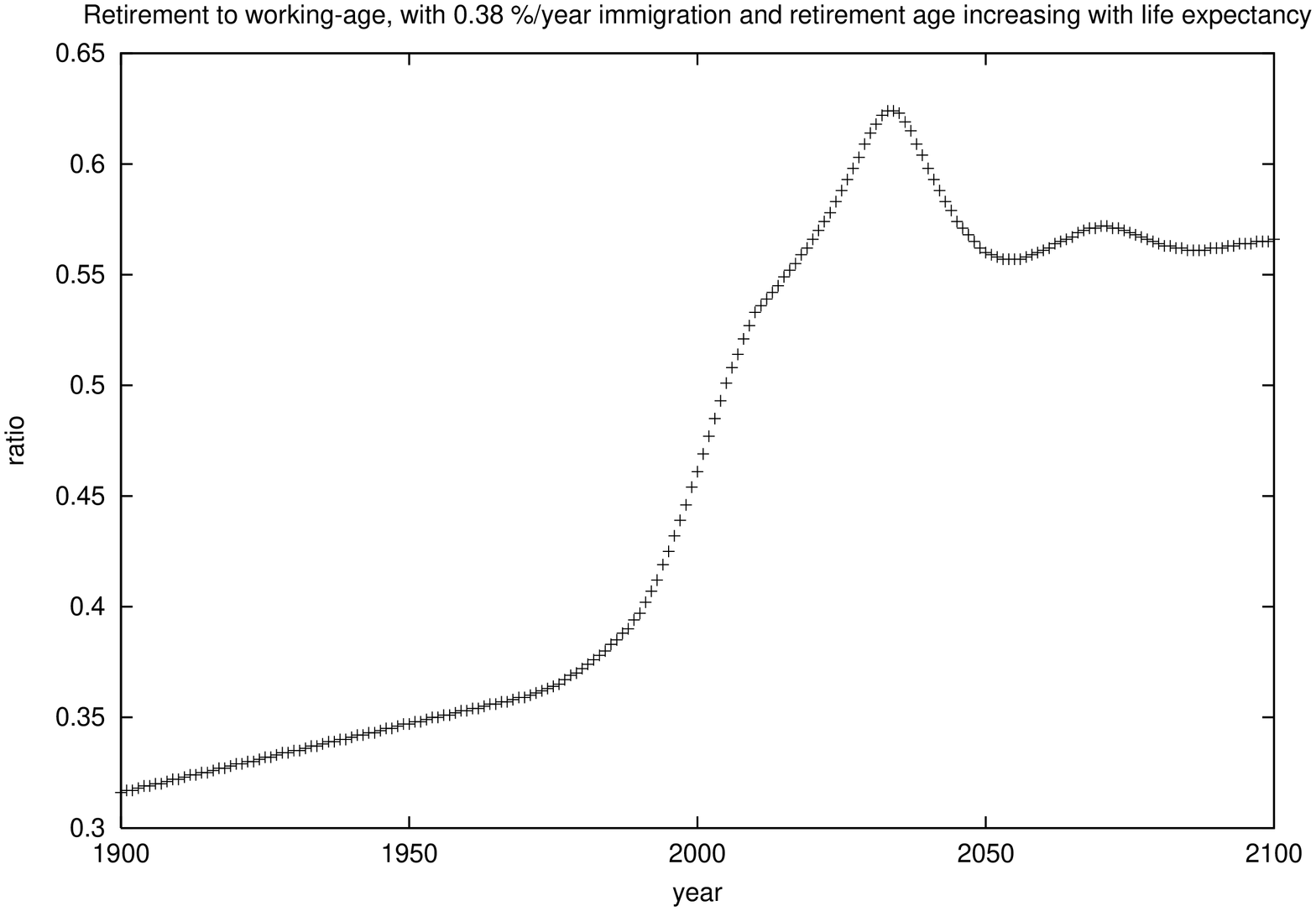}
\caption{Ratio of people above retirement age, to working-age people older than
20 years, as predicted in Ref.~6 using extrapolated Gompertz laws.}
\end{figure}

\section{Demography}

The mortality rate $q$ is the fraction of people of age $x$ who die within the 
next time unit, i.e. before they reach age $x+1: \quad q(x)=[S(x+1)-S(x)]/S(x)$.
Here, $S(x)$ is the probability to survive from birth to age $x$. This quantity
$q$ can by definition not increase beyond $q=1$. It depends on the time unit
which is typically a year for humans and a day for flies and worms. A better
quantity, which can increase beyond unity, is the derivative $\mu$ for 
infinitely small time steps instead of discrete time steps, 
$$ \mu(x) = - d \ln S(x)/dx \quad ,\eqno(1)$$
called here the mortality function (also denoted as hazard rate or force of 
mortality \cite{thatcher}). If life tables with yearly units are published, 
then $\mu$ can be approximated through
$$ \mu(x + 1/2) \simeq \ln S(x) - \ln S(x+1)) \eqno (2)$$
which also can increase beyond unity. The astronomer Halley tried about three centuries
ago to find some laws governing human mortality, but only in the 19th century
Gompertz found the law which is valid when childhood diseases are overcome, 
Fig.~1:
$$ \mu \propto \exp(bx) \eqno(3)$$
with an empirical parameter $b$ increasing for humans over the centuries. 
It holds also approximately for many animals in protected environments like 
zoos and laboratories \cite{vaupel}, while in the wild many animals are eaten
by predators before they reach their genetically possible age.
Below the age of 25 years for humans, deviations are seen: The mortality (rate
or function) is high at birth (most human embryos die before birth), then sinks
to a broad minimum between 5 and 12 years, and only then increases monotonically
up to old age. 

For humans, the last two centuries have seen a doubling of the life expectancy
at birth, $\int S(x) dx = \int x \mu(x) S(x) dx$, from 40 to 80 years. Figure 2 
shows, for Swedish women since 1750, both the prolongation of life and the 
increasing reliability. (We did not all die at age 40 in earlier 
centuries. If half of the babies die in their first year, and the other
half lives until 80, then the life expectancy is 40 years.) We see clearly that 
the life expectancy no longer increases as fast as in the first half of the 
20th century, but the present lower rate of increase showed no sign of further
reduction during the last decades. Figure 3 shows that this increase of life 
expectancy comes not only from the reduction of child mortality but also 
from an increase of the remaining life expectancy at age 65.

Much more recent is the reduction of the birth rate (number of children born per
women during her lifetime) below the replacement level of about 2.1. The two 
German states started this around 1970, due to "the pill", and in West Germany 
the birth rate scattered about 1.4 in the last three decades. In France it is 
higher, in Spain and Italy lower. World War II was started by Nazi Germany with
the excuse of "Volk ohne Raum", than the Germans needed more living space; so
a reduction of the native population (enlarged by immigration) did not seem bad
around 1970. In the meantime, however, the reduction in the number of young 
people coupled with the increase in the number of old people is seen as a threat
to the usual way in which you should finance my retirement. If the strongest
age cohort in the year 2030 will be people at age 70, we can hardly afford
an average retirement (healthy Germans) at 62. 

Thus we \cite{martins} (and others) predicted the future ratio of 
pensioners to working-age people, assuming that after the year 2011 the 
retirement age is increased from 62 years by 0.6 years for every year by which 
the life 
expectancy at birth increases, and that starting in 2005 an immigration of
0.38 percent per year (of the population) of people aged 6 to 40 years keeps
the total population stable. We see that the dangerous peak around the year
2030 is followed by a plateau in this ratio. (Working was assumed to start at 
age 20.)

In a comparison of life tables for different countries and different countries,
a certain degree of universality was found for the human Gompertz law 
$\mu \propto \exp(bx)$: The mortality for centenarians was about the same
\cite{mildvan,gavrilov,azbel1}. Thus 
$$\mu \simeq 7 b \cdot \exp[b(x-X)] \eqno (4) $$ 
with a 
characteristic age of $X \simeq 103$ years for the whole human species, while
$b \sim 0.1$ increases with time. 
(For $x < X$ the differences between $q$ and $\mu$ are quite small.) Moreover,
Azbel \cite{azbel2} found (with some deviations) a universality even for the
younger ages where Gompertz is invalid, Figure 1. He found the mortality 
$q_x(c,t)$ at age $x$
(for country $c$ and calendar year $t$) to be a universal function $f_x$ of 
infant mortality $q_0 = q_{x=0}(c,t)$ and age $x$: $q_x(c,t) = f_x(q_0)$. 
The function $f_x$ no longer depends explicitly on $c$ and $t$, in contrast to
$q_x$. 
Thus if country A has at present a known mortality
function of age, then another country B roughly has the same mortality function
if we change the calendar year $t$ such that the infant mortality in B at time
$t$ agrees with the present infant mortality in A. If the Gompertz law would
be valid for all ages, Azbel's universality \cite{azbel2} would already follow 
from Eq.(4) since it contains only one free parameter $b$ for all human 
societies. These universality laws suggest that
extrapolations like Fig.~4 may, with some shift in time $t$, be valid also
for developing countries, if they do not take early action to keep the birth
rate near the replacement level of 2.1 or whatever else is needed to offset
deaths and net emigration.

A decade ago, mortality maxima were observed \cite{carey,curtsinger} for 
flies. Have they found the fountain of youth such that we get healthier again
with increasing age ? Humans at least, Fig.~1, do not show such maxima in 
reliable statistics, though USA data published in the 1990 showed them. 
(Reliability seems to increase from USA to Western Europe and from there to 
Sweden.) Perhaps above 110 years human mortality reaches a plateau 
\cite{thatcher,robine}.
But a comparison of Figures 3 and 4 in \cite{robine} shows that for the more
reliable half of the European data, the highest claimed ages were appreciably
below those of all the data: The more reliable the data are the smaller are
the deviations from the Gompertz law. Perhaps for the oldest old the mortality 
still increases with age, but only linearly \cite{perls} and not exponentially:
Neither acceleration nor deceleration of mortality. More arguments against 
mortality deceleration for humans are given elsewhere \cite {zanjan,dresden}.

While at present we thus should be cautious about buying fountains of youth for 
humans, future decades might produce genetically modified humans with longer
life expectancies. The little worm {\it Caenorhabditis elegans} survives bad
times (no food, ...) by reducing all life functions during a "dauer" state, in
agreement with computer simulations \cite{heumann}. Even flies and some mammals
live longer if put on a starvation diet \cite{mair}. But do we want to live longer if the gained
life span is spent in a coma, or in hunger? More attractive is a very old 
elixir of youth, red wine. According to \cite{howitz}, the polyphenol 
resveratrol in red wine activates the Sir2 gene and lets yeast cells life
70 \% longer. Indeed, Jeanne Calment is widely (though not universally) believed
to be the oldest human being and died in 1997 at age 122 in Southern France,
having drunk red wine moderately. Let me see if drinking it beyond moderation lets
me beat her world record.

\section{Why do we age?}

It may be an exaggeration that there as many theories of ageing as there are 
ageing theorists, but nevertheless we have lots of theories. They even might
all be correct, since ageing may have many causes. Also, some theories do not
exclude each other, describing only different aspects of the same phenomenon. 

120 years ago, Weismann suggested that we die to make place for our children
\cite{weismann}.
This is very altruistic but also very unrealistic, since of two different 
races of the same species the one which produces more children will win the
Darwinian struggle of survival of the fittest. Those who live longer at 
otherwise unchanged parameters produce more children and thus win in the short 
term even if on long time scales they drain the environmental resources
stronger and might finally destroy the ecosystem, including themselves. (If,
however, longer lifetime is coupled to lower birth rates, the Weismann idea
becomes viable as explained later.) 

Medawar \cite{medawar} suggested more than half a century ago the still 
relevant mutation accumulation theory: Bad mutations killing us in young
age before we get children will die out since they are not given on to the
offspring; bad mutations killing us after we produced many children are
given on to future generations. Thus after some time the population should 
contain few hereditary diseases affecting us in young age, but many affecting 
us in old age. Thus the probability to die from them increases with increasing 
age. 

A now widespread idea are oxygen radicals created by metabolism and destroying
the DNA, carrier of heredity, during our life. This theory is not necessarily
in contradiction with the mutation accumulation; instead it is a biochemical
explanation for these mutations. 

Telomeres are sections at the end of the DNA which are lost at every 
duplication of the cell. If the number of telomeres in this way has become too
small, the cell stops dividing: Hayflick limit. A recent ageing theory
\cite{aviv} is based on these telomeres, and perhaps at the time of the 
conference I can present more simulations. 

Longevity genes would prolong life, have been found to work for many animals,
and are perhaps connected with the red wine effect mentioned at the end of
section 2. Again, their existence does not contradict the other theories: A
longevity gene may produce more telomerase, an enzyme which restores lost 
telomeres. Or it may enhance scavenging of dangerous oxygen radicals and thus
reduce the mutation rate relevant for mutation accumulation.

Reliability theory \cite{gavrilov} may work also for the ageing of 
automobiles and was connected to the Penna model (see below) in \cite{pletcher},
see also \cite{cebrat4}. It assumes the organism or car to consist of $m$
irreplaceable blocks; failure of one of these blocks causes the whole system to 
fail. Each block consists of many equivalent elements; the initial numbers of 
properly working elements within a block follow a Poisson distribution.  A block
fails if all its elements fail; each element ages with a constant failure rate 
$1/X$. Then Gavrilov and Gavrilova \cite{gavrilov} recovered analytically
the Gompertz law, Eqs.(3,4), for age $x \ll X$ and a mortality plateau $\mu =
m/X$ for very high age $x \gg X$. The characteristic age $X$ in Eq.(4),
valid for all humans, then is the average lifetime of the single elements.

The following sections will report computer simulations of the mutation
accumulation idea. 

\section {Simulations of mutation accumulation}  

This section restricts itself to those individual-based ageing models which were
investigated in papers of different groups. Historically the first are
those of Partridge-Barton type \cite{partridge}, followed by the most widely 
used Penna model \cite{penna}, while the Stauffer model \cite{dresden} is more
of conceptual than of practical value but therefore forms our starting
point.

In contrast to Weismann, we do not die to make place for our children. But if we
fix the number of children, then the idea \cite{dresden} works: The birth rate 
(per iteration) is assumed to be inversely proportional to the time between
the minimum age of reproduction, $x_m$, and the genetic death age, $x_d$. 
Hereditary mutations accumulate over the generations, and each may independently
change both characteristic ages $x_m$ and $x_d$ by one time unit. Individuals
may die before Thai genetic death age from hunger etc, which is taken into 
account by a Verhulst death probability proportional to the population size,
as in a logistic equation. Then automatically a reasonable distribution of 
death ages emerges and death is explained as coming from random mutations 
plus a trade-off between longevity and high birth rate. The catastrophic 
senescence of Pacific salmon, the death of Northern cod though over-fishing,
the minimum population size for social animals, and the emergence of female
menopause were simulated successfully \cite{ortmanns,radomski,sousamenop}.
However, in general  \cite{makowiec}
the mortality increases linearly with age, instead of the desired exponential
Eq.(3). Also, the minimum age of reproduction is distributed among unrealistic
short ages, even in a much more complicated model of a whole ecosystem 
\cite{chowdhury}.

This trade-off between longevity and high birth rate is mentioned a lot in the
biological literature. The most direct but not the only way to realize it 
genetically are mutations with antagonistic pleiotropy \cite{partridge,pleio}: 
these genes have positive effects in youth and negative ones at old age. 

\bigskip
Computer simulations of ageing started by putting fluctuations into the
phenomenological model of Partridge and Barton 
\cite{partridge,ray,jan,dasgupta,heumann,onody,sousasex}. 
Originally it assumed only three ages zero, one and 
two, with a juvenile survival rate $J$ from zero to one, and an adult survival
rate $A$ from one to two. It was first thought to give unrealistic mortality
functions and difficulties if generalized to many age levels, but \cite{onody}
repaired this by slight modifications, and \cite{sousasex} included sexual 
reproduction in it. But the lack of an explicit genome makes it less attractive
than the Penna model described now.
\bigskip

The Penna model is by far the most widespread method to simulate biological 
ageing. Most of the literature up to 1998 is cited in \cite{book}, and later
work up to 2000 in \cite{dresden} for asexual and \cite{anais} for sexual 
reproduction. The genome is represented by a bit-string ($10^1 \dots 10^3$ bits
were simulated) giving bad mutations. A bit set to zero
is healthy, a bit set to one means that a hereditary disease starts to reduce
the health at the age to which the bit position corresponds. The first bits
describe diseases starting in youth, which are rare, and the latest bits 
correspond to the much more frequent diseases at old age. Three active diseases
kill, and so does a Verhulst death probability proportional to the total
population size. Each time interval, every individual above the minimum age of 
reproduction produces offspring which differs from the parent by a random 
mutation of the bit-string genome. In the sexual version with recombination of
the two bit-strings of the genome, dominant mutations affect the health already
if only one bit-string is mutated, while recessive mutations become dangerous 
only if both bit-strings are mutated. Bigamy with three bit-strings was discussed
in \cite{triple}.

\begin{figure}[t]
%\figurebox{22pc}{15pc}{}
%\epsfbox{vancouver5.eps}
\includegraphics[angle=-90,scale=0.5]{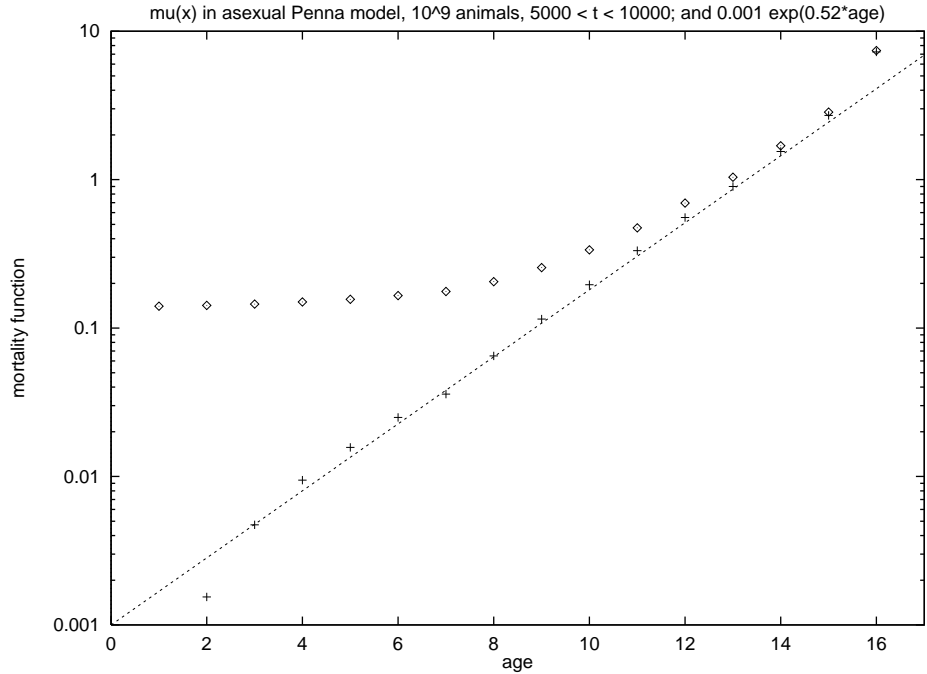}
\caption{ Mortality function $\mu$ for the standard asexual Penna model, with
(upper data) and without (lower data) the deaths from the Verhulst factor.}
\end{figure}

\begin{figure}[t]
%\figurebox{22pc}{15pc}{}
%\epsfbox{vancouver6.eps}
\includegraphics[angle=-90,scale=0.5]{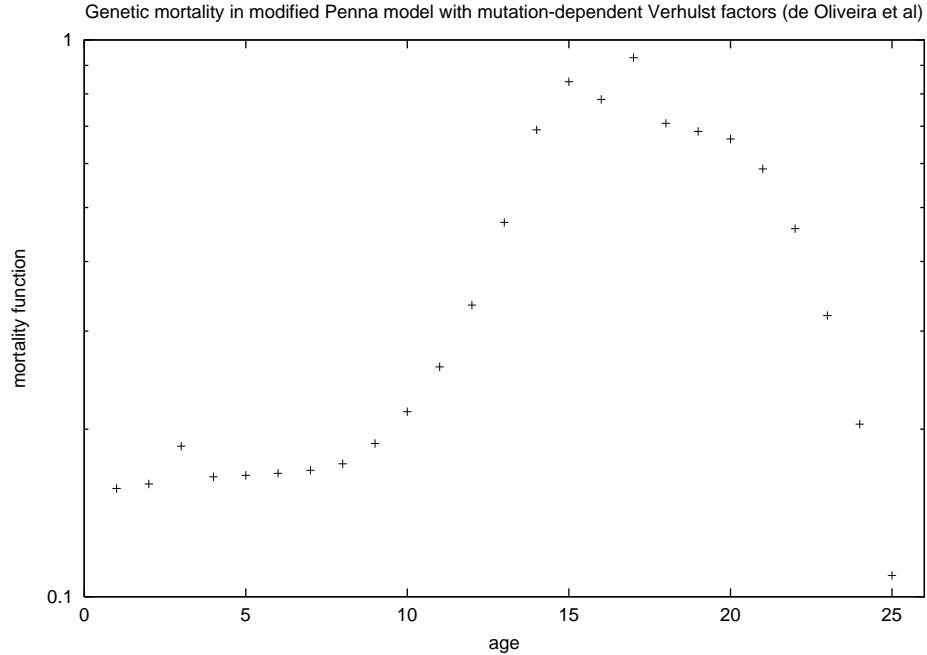}
\caption{Semilogarithmic plot of mortality function in a modified Penna model}
\end{figure}

Figure 5 from \cite{book} shows the resulting mortalities, with and without the
deaths from starvation or lack of space. The purely genetic deaths follow nicely
the exponential Gompertz law, except for the youngest and oldest ages. The
limit of exactly three mutations killing can be softened \cite{coe} to give 
slight downward deviations from the Gompertz law as in Swedish mortalities, 
Fig.~1, or a mortality plateau as claimed in \cite{robine}. It can also be 
abolished in favour of Verhulst factors depending on the number of active
mutations; then a mortality maximum even more pronounced than for flies
\cite{carey,curtsinger} is obtained in Fig.~6 from \cite{moss}. Simulations
of biologists, in contrast, could not yet get such mortality maximum 
\cite{mueller}. Pacific salmon, Northern cod, and Alaskan wolves were 
simulated successfully long ago \cite{book}, and Lyapunov exponents \cite{atb},
Brazilian lobsters \cite{lobster}, child mortality 
\cite{cebrat2,magdon}, prey-predator relations on lattice \cite{he},
and speciation more recently \cite{cebrat,luz}. Particularly relevant for our
section 2 are the Penna model simulations of the demographic changes in the
20th century \cite{cebrat3}. The mortalities do not change much \cite{alle}
if the genome may contain the same gene in several copies called 
"paralogs" \cite{cebrat4}.

This section ends with a technical warning: If the population is kept
constant artificially, as is tradition in theoretical biology, instead of being 
allowed to fluctuate as in nature, then the results are only qualitatively, not
quantitatively, the same \cite{const}.

\section{Sex}
Sexual reproduction was introduced into the Penna model long ago \cite{book}, 
for the Partridge-Barton type \cite{sousasex} and the Stauffer model 
\cite{sousamenop} only recently. Even bacteria exchange genome via "parasex"
\cite{bactsex}, 
and computer simulations with the Penna model showed this parasex to give
fitter individuals than pure asexual cloning \cite{parasex}. These simulations 
included ageing of bacteria, as found later experimentally \cite{bacteria}.
Less clear is the 
need for males in species with two sets of chromosomes, from father and mother
\cite{anais}. Only with some effort \cite{martins2} could feeding the males be 
justified in the Penna model; and no simulation yet showed hermaphroditism
to be by far the fittest way of reproduction. On the other hand, sexual 
reproduction is clearly preferable as a protection against parasites or 
environmental catastrophes \cite{anais}. So, the sex wars can continue
\cite{sexwar}.

Menopause or it's analog is the cessation of female reproductive power at middle
age. In spite of widespread prejudice, it is not restricted to humans (and
pilot whales) but even occurs in some flies \cite{fly}. Computer simulations
showed, without any specific human assumptions like tradition of knowledge,
that menopause can emerge automatically \cite{meno,sousamenop}, provided the 
risk for the mother of giving birth increases with increasing age and/or the
child depends on the mother for its initial survival. Are men needed for 
survival of the children? Only indirectly \cite{book}: without them women would
follow Pacific salmon and die rapidly after giving birth; the lack of a male
analog for a sharp menopause makes males useful for producing children even at 
older age, thus prevented evolution to kill females after their cessation of 
reproduction.

\section{Conclusion}

Computer simulation of mutation accumulation models has advanced a lot in one
decade and has applications like retirement rules. Particularly important seem
the menopause explanations \cite{meno,sousamenop} showing
that such effects are not restricted to humans. Simulations of alternative 
theories of biological ageing \cite{aviv} are mostly lacking.

\section*{Acknowledgements}
I thank N. Jan, S. Moss e Oliveira, P.M.C. de Oliveira, T.J.P.Penna, A.T. 
Bernardes, S. Cebrat. J.S. S\'a Martins, A.O. Sousa and many others for
fruitful collaboration over many years.

\end{document}